\documentclass[12pt,a4paper]{article}
\usepackage{a4wide}
\begin{document}

\thispagestyle{empty}

\begin{flushright}
  THEF-NYM-98.02 \\
\end{flushright}

\vspace{2\baselineskip}

\begin{center}
  {\large\bf Comment on ``$\pi\!N\!N$ Coupling \\
  from High Precision np Charge Exchange at 162 MeV''}
  \\[1.5cm]
  M.C.M.\ Rentmeester, R.A.M.\ Klomp, and J.J.\ de Swart,
  \\ [0.5cm]
  {\it Institute for Theoretical Physics,  University of Nijmegen, \\
    Nijmegen, The Netherlands}
\end{center}

\vspace{1cm}

\begin{center}
  Submitted to Phys.Rev.Letters on March 29, 1996, \\
  Published in Phys.Rev.Letters {\bf 81}, 5253 (1998)
\end{center}

\vspace{1cm}

\begin{center}
  ABSTRACT\\
\end{center}

In this updated and expanded version of our delayed Comment we show that 
the np backward cross section, as presented by the Uppsala group\cite{1}, 
is seriously flawed (more than 25 sd.). The main reason is the incorrect
normalization of the data.
We also show that their extrapolation method, used to
determine the charged $\pi\!N\!N$ coupling constant, is a factor of
about 10 less accurate than claimed by Ericson {\it et al.}.
The large extrapolation error makes the determination of the
coupling constant by the Uppsala group totally uninteresting. 

\vfill

E-mail: swart@sci.kun.nl

\newpage
\pagenumbering{arabic}

In a not so recent anymore Letter\cite{1}, a measurement of
the np differential cross-section in the backward
direction at a single energy T$_{\rm lab}$=162 MeV
was reported.
These 31 data were then used to extract for
the charged pion-nucleon coupling constant the large value
$f_{c}^{2}=0.0808$.
An incredibly small extrapolation error of
only $0.0003$ and a normalization error of $0.0017$ are claimed.
The systematic errors, however, have not been properly dealt with.
This Uppsala value for the charged coupling constant is in agreement
with the old-fashioned textbook values, but
in strong disagreement with modern determinations\cite{vanc}.
\\[\baselineskip]
First we make the observation that this coupling constant
has been determined in the last decade by several groups
(and not only the Nijmegen group as suggested in the Letter \cite{1})
in various energy-dependent Partial Wave Analyses (PWA's).
For a review see reference \cite{vanc}. These PWA's
give very good fits to about 12,000
np \cite{2,3,4}, $\pi$N \cite{6,7}, and
$\overline{{\mathrm p}}$p charge-exchange \cite{5} scattering
data.
The values of $f_{c}^{2}$ determined in these different
energy-dependent PWA's from thousands of data are all in excellent
agreement with each other, and in flagrant disagreement with the
determination \cite{1} from the merely 31 Uppsala data.
A representative value (with error) for this coupling
constant is $f_{c}^{2}=0.0748(3)$ \cite{8}.
This charged coupling constant $f_c$ is via charge independence
related to the pp$\pi^0$ coupling constant $f_p$.
This latter coupling constant has been determined from almost two
thousand pp data \cite{vanc} with the result $f_p^2=0.0753(5)$.
An incredibly large breaking of charge independence would therefore
be implied if the Uppsala value of the charged coupling constant would
be correct.
\\[\baselineskip]
The backward np differential cross section is sensitive to
$f_{c}^{2}$. That it is {\em therefore} a good place to
determine this coupling constant is a widespread
misunderstanding. This has been shown \cite{8} in an
energy-dependent PWA of the np data.
The backward np data do not show any {\em particular}
sensitivity to $f_{c}^{2}$.
In table V of \cite{8} one can see that in our energy-dependent
PWA using {\em all} np scattering data and {\em all} types of
observables, $f_{c}^{2}$ shows no special sensitivity to
any particular type of observable.
The claim in \cite{loi} that ``the Uppsala group has shown the contrary
using pseudodata'' is false. The Uppsala group did not study anything
else but differential cross sections. Therefore they cannot 
make any statement about the {\em relative} importance of various
observables (like differential cross sections, polarizations, spin
transfer coefficients, etc.) in the determination of the coupling
constant.

No particular sensitivity for any particular observable implies
that all datatypes contribute with about the same weight.
This means that the statistical errors in the different analyses
are roughly inversely proportional
to the square root of the number of datapoints.
The pole extrapolations use about a factor of 100 less
data than the energy dependent PWA's,
implying a statistical extrapolation error that is about 10 times
larger than the error in the PWA's. Such a large error makes the
determination of $f_c^2$, as described in the Letter \cite{1}, totally
uninteresting. It must be clear from statistical reasons that
a rather small data set {\em cannot} be used for an {\em accurate}
determination.

In the same paper \cite{8} it has been explicitly shown,
using physical extrapolation techniques, that analyzing
backward np data at a single energy, as in ref. \cite{1},
gives values of $f_{c}^{2}$ with a large spread that results
in a total error of $0.003$, which is 10 times larger than the
extrapolation error claimed in \cite{1}.
This was confirmed by Arndt {\it et al.} \cite{9},
who used exactly the same techniques as used in \cite{1}
for {\em all} the available backward data, and not for only one
dataset as was done in \cite{1}.
Their values for $f_{c}^{2}$ as determined at a single energy
vary from 0.061 to 0.091 with an average of $0.075$ and an error
of $0.009$, which is 30 times the extrapolation error
quoted in \cite{1}.

In their Reply to our Comment the authors of
the Letter \cite{1} imply that ``the analysis of Arndt
{\it et al.} \cite{9} is not detailed enough and their examination
of the input data not critical enough''. That expresses exactly
our opinion about the work of Ericson {\it et al.} as presented
in \cite{1} and subsequent publications.
{}From decades of experience with the work of Arndt we know
that it definitely does not apply to the work of Arndt.

In the Letter \cite{1} the authors use a self-invented
extrapolation method, which they call the Difference Method.
However, they did not study properly the
systematic errors in their new method.
This was done in \cite{vanc}, where it was shown
that the model-dependence of their method is enormous.
This large model-dependence gives rise to
very large systematic errors in their value for the coupling constant
and in their estimate of the error.

The authors of the Letter state in the Abstract that they can
extrapolate precisely and model-independently to the pion pole.
That is definitely incorrect.
Their extrapolation method is strongly model-dependent,
with large systematic errors,
and as inaccurate as any other extrapolation method.
Not better, not worse.
A new extrapolation method that {\em really} 
produces extrapolation errors a factor of 10 smaller than
other extrapolation methods would have been a sensational
discovery in numerical analysis and/or statistics.
This Difference Method is definitely not {\em better} than the
standard Chew extrapolation technique. However, it is certainly much
more cumbersome. It is so cumbersome, that the Uppsala group could
not properly determine their errors.

The pole-extrapolation method used by Ericson {\it et al.}\ relies
heavily on the absolute normalization of the data.
Normalizing np cross-sections is very difficult.
In their determination of $f_{c}^{2}$ the normalization is another
important source of uncertainty.
In energy-dependent PWA's, as in \cite{2},
one does not need normalized data to determine the coupling
constant; one can use the {\em shapes} of the measured
differential cross-sections.

Do not misinterpret the above statement. We do not say that we
apply all our methods directly to unnormalized data.
We normalize data very accurately with the help of our PWA.
This has been explained extensively in most of our
publications\cite{8} about PWA's.
This is definitely one of the successes of energy-dependent PWA's;
we determine the normalization of differential cross sections in
np scattering with a typical uncertainty of about 0.5 \%.
This is a lot better than the 4 \% normalization error used in the
Letter.
The remarks in the Reply about ``loose normalizations'' show an
unfortunate lack of knowledge by the Uppsala group of the methods of
modern PWA. The corresponding sentences in their Reply 
do not correspond to the truth, but are
fabrications of the unbridled fantasies of the authors of the Reply.
\\[\baselineskip]
The authors have applied their method for extraction of $f_{c}^2$
to data which cannot be described satisfactory by either the
Nijmegen PWA \cite{2} or the VZ40 PWA of Arndt {\it et al.}\ \cite{4}.
The Nijmegen PWA gives, after refitting, $\chi^{2}= 264.0$
for these 31 data and the VPI\&SU PWA \cite{4} gives $\chi^{2}=236.7$.
One reason for the bad fit can be seen in the large discrepancy
between the shape of the newly reported data and the shape of the
older data of Bonner {\it et al.}\ \cite{b} at exactly the same energy.
The authors should have reported $f_{c}^{2}$ from applying their
extrapolation method to the Bonner data and compared the results.

In their Reply the authors claim:
``the data of the present experiment are of a far better quality than
those of Bonner at 162 MeV''.
When reading this above quote one must realize here that people, not
known for their familiarity with data analysis, are claiming that
their own experiment is the best. This is definitely not the
opinion of the Nijmegen data analysis group. They find in their
careful, detailed, and critical analysis of the data that the
old Bonner data are of better quality than the new Uppsala data.
The Bonner data are included in the Nijmegen and the VPI\&SU
databases; the Uppsala data are not!

The new data disagree not only with the Bonner data,
they disagree with the whole Nijmegen np data set, currently
consisting of circa 5000 data below 500 MeV.
They disagree, because of their wrong shape.
The shape of the Uppsala differential cross section is
{\em more than} 25 sd away from both the Nijmegen and the VPI\&SU databases.
More than 3 sd is already called ``wrong''.
\\[\baselineskip]
We \cite{mart} have studied these data to see what is really wrong
with them.
In their experiment the Uppsala group performed 3 different measurements
in 3 angular regions, which were then separately
normalized. They have 49 (partially overlapping) datapoints.
These 3 datasets are then combined to one dataset with 31 points.
We have pinpointed two errors.
Firstly, in those angular regions where these datasets overlap one
clearly notices internal inconsistencies in the slopes.
This discrepancy is nowhere \cite{ol} discussed in the Uppsala papers,
but just ignored.
Secondly, we can improve the $\chi^2$ for the total dataset
dramatically by just renormalizing these 3 sets and
discarding 4 datapoints \cite{mart}.
However, we cannot improve them so much that the data become acceptable.
They are, after renormalization and discarding the 4 bad points,
still more than 3 sd. away from the Nijmegen database.
This is for statistical reasons unacceptable.
But ..... , we made the Uppsala data at least {\em almost} acceptable.

In their Reply the authors refer to the H\"{u}rster {\it et al.}
data \cite{H}, which are not used (but intensively studied) in the
Nijmegen PWA's and also not used in the Arndt {\it et al.} PWA's
because of their high $\chi^2$.
The authors claim that the shape of the incorrectly normalized Uppsala
data agrees with the shape of these H\"{u}rster {\it et al.} data.
This is almost certainly incorrect.
Because then we need to assume that the Freiburg people made exactly
the same mistakes as the Uppsala people in normalizing their data and
that these data have the same kind of internal inconsistency.
We see no reason to make such drastic assumptions.
Also the $\chi^2$/datapoint for the H\"{u}rster {\it et al.} data
is much smaller than for the (incorrectly normalized) Uppsala data.\\

The authors state in their Reply that possibly the inclusion of the
Bonner data in the Nijmegen and VPI\&SU PWA's is responsible for the
large $\chi^2$ of the Uppsala data. This also is incorrect.
We have done PWA's in which we discarded all Bonner data;
the $\chi^2$ of the Uppsala data was still unacceptably high.
These and other studies performed in Nijmegen show that the Uppsala
data are in disagreement with the whole
database and {\em not} only with the Bonner data.
\\[\baselineskip]
In their Reply to our Comment it is stated that: ``Their Letter argues
that a rather small, but well-controlled data set on a relevant
observable can be used for an accurate determination when carefully
analyzed.'' We observe that the dataset is indeed rather small,
only 31 points. This is insufficient for an {\em accurate}
determination.
According to the energy-dependent PWA's the dataset is definitely
{\em not} well-controlled.
In 1993 it was already shown in \cite{8} that the backward differential
cross section, combined with a pole-extrapolation method,
is not an {\em especially} relevant observable. In the Reply one can read
the unbelievable remark that our proof is not relevant to their approach.
We find this an unprofessional way of discarding unwanted facts.
We think that the statement ``when carefully analyzed'' is neither
applicable to these incorrectly normalized data with internal
inconsistencies, nor to their extrapolation analysis with their unnoticed, 
huge, systematic errors.
\\[\baselineskip]
Our conclusions are:
\\
i) The experimental data {\em as presented} are seriously
flawed (more than 25 sd.).
This is mainly caused by the way these data are normalized.
Similar data\cite{u} at
96 MeV from the same group are not included in the Nijmegen
database\cite{2} either
because they too disagree significantly with the total
dataset.
\\
ii) Achieving an {\em accurate} determination of $f_{c}^{2}$ from
the backward np data at one single energy is a rather
unrealistic exercise.
The label ``dedicated'' for such experiments is presumptuous and
completely unwarranted.
We have shown that to determine $f_{c}^{2}$ accurately
the energy-dependent PWA's are vastly superior over the
pole-extrapolation methods.
\\[\baselineskip]
We would like to thank Prof. N. Olsson for providing us the data, and
R. Timmermans and T. Rijken for stimulating discussions.

\end{document}